\documentclass[aps,pra,floatfix,twocolumn,tightenlines]{revtex4-2}
\usepackage{amsmath}
\usepackage{braket}
\usepackage{bbold}
\usepackage{latexsym}
\usepackage{graphicx}
\usepackage{color}
\usepackage{xcolor}
\usepackage{amsfonts}
\usepackage{mathtools}
\usepackage{mathrsfs}
\usepackage{soul}
\usepackage{pythonhighlight}

\begin{document}

\title{Enhanced resolution chirped-pulse interferometry}

\author{M. M. B\'{e}rub\'{e}$^1$, M. D. Mazurek$^{2,3}$, and K. J. Resch$^1$}
\affiliation{$^1$Department of Physics \& Astronomy and Institute for Quantum Computing, University of Waterloo, 200 University Ave.~W, Waterloo, Canada N2L 3G1\\ $^2$Department of Physics, University of Colorado Boulder, Colorado 80309, USA \\
$^3$Associate of the National Institute of Standards and Technology, Boulder, Colorado 80305, USA }

\begin{abstract}
Chirped-pulse interferometry (CPI) is a classical low-coherence interferometry technique with automatic dispersion cancellation and improved resolution over white-light interference.  Previous work has shown that CPI with linearly-chirped Gaussian laser pulses achieves a $\sqrt{2}$ improvement in resolution over conventional white-light interferometry, but this is less than the factor of 2 improvement exhibited by a comparable quantum technique.  In this work, we show how a particular class of \emph{nonlinearly}-chirped laser pulses can meet, and even exceed, the factor of 2 improvement resolution.  This enhanced resolution CPI removes the remaining advantage of quantum interferometers in dispersion-cancelled interferometry.
\end{abstract}

\maketitle

\section{Introduction}

Chirped-pulse interferometry (CPI) \cite{kaltenbaek08} is a low-coherence interferometry technique, a class of technologies with important applications in biomedical imaging and advanced manufacturing for noninvasive imaging \cite{fujimoto95}, surface topography \cite{degroot15}, and precision drilling \cite{webster11}.  
CPI \cite{kaltenbaek08} aims to achieve all the metrological advantages of the quantum, Hong-Ou-Mandel (HOM), interferometer \cite{hong87} without the inherent difficulties of working with single photons.  These advantages include automatic cancellation of even orders of unbalanced chromatic dispersion \cite{steinberg92} and a resolution enhancement over a white-light interferometer \cite{abouraddy02}.  CPI can achieve these features while producing signal large enough to measure with standard photodiodes, instead of single-photon detection and coincidence measurements \cite{kaltenbaek08}.  

\begin{figure}[b]
\begin{center}
\includegraphics[width=0.8\columnwidth]{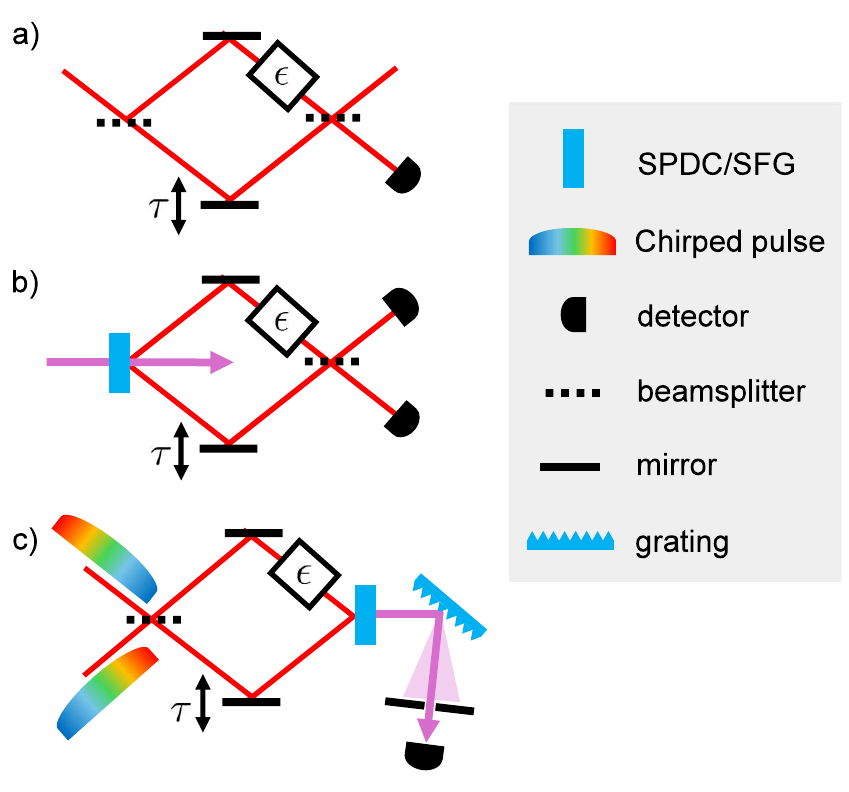}
\end{center}
\caption{Three different types of low-coherence interferometer: a) white-light Mach-Zehnder interferometer, b) Hong-Ou-Mandel interferometer, c) chirped-pulse interferometer.  In all cases, we will consider the time-domain interference signal as a function of optical delay, $\tau$ and include the possibility of some unbalanced second-order dispersion characterized by a strength, $\epsilon$. For the white-light interferometer, broadband light is split at a beamsplitter, travels along two different arms, and is recombined at a second beamsplitter.  The output intensity in one of the final paths is measured using a standard detector.  For the Hong-Ou-Mandel interferometer, energy-time entangled photon pairs are generated using, for example, spontaneous parametric downconversion.  Each photon travels one arm before the pair is recombined on a beamsplitter and detected using a pair of single-photon counting detectors.  In the case of CPI, a pair of synchronized oppositely chirped pulses are combined on a beamsplitter, then the light travels through two arms and drives sum-frequency generation.  The intensity of the output light is detected in a narrow spectral range.
\label{fig:interferometersfigure}}
\end{figure}

Low-coherence interferometers specialize in position measurements where the axial resolution is limited by the coherence length of the light source, the inverse of the optical bandwidth.  In this paper, we discuss the three types of time-domain interferometers shown in Fig.~1: a white-light interferometer (WLI) Fig.~\ref{fig:interferometersfigure}a), the Hong-Ou-Mandel (HOM) interferometer Fig.~\ref{fig:interferometersfigure}b), and CPI Fig.~\ref{fig:interferometersfigure}c).  All three types of interferometer measure the interference between a reference arm with a variable optical delay, $\tau$. WLI is based on linear, or single-photon, interference with a broadband light source, HOM interferometry on two-photon interference of energy-time entangled photon pairs, and
CPI is based on an optical cross-correlator fed by two oppositely chirped laser pulses.  In this work, we will consider the impact on the interference signal due to unbalanced chromatic quadratic dispersion, $\epsilon$, in one of the arms. Of the three types of interferometers, the WLI is most widespread but it is sensitive to the dispersion, lowering both resolution and interference visibility.  In contrast, HOM and CPI are examples of automatic even-order dispersion cancellation techniques \cite{steinberg92}---  both maintain sharp interference features even with uncompensated dispersion---and have higher intrinsic resolution than WLI \cite{resch07, lavoie09, mazurek13, abouraddy02, nasr03}.  Note that automatic dispersion cancellation should more accurately be referred to as automatic dispersion \emph{reduction} once physically-realizable frequency correlations or entanglement are considered \cite{resch09}.  

\emph{Quantum}-optical coherence tomography \cite{abouraddy02, nasr03} was developed around HOM interferometry to take advantage of automatic dispersion cancellation and the resolution enhancement for biomedical imaging. The resolution enhancement refers to the fact that, for photon pair sources with Gaussian spectra, the width of the HOM dip can be a factor of 2 narrower than the width of the white-light interference pattern generated by one beam from the pair source \cite{nasr03}.  In order to achieve this factor of 2 resolution enhancement, the HOM interferometer requires very strong energy-time entangled photon pairs with no additional spectral filtering between the source and detection; if the photon pairs are in a product state with Gaussian spectra, one only expects a $\sqrt{2}$ resolution enhancement and no dispersion cancellation \cite{abouraddy02, nasr03}.  Comparing the width of the CPI signal to that of the white-light signal generated by one of the laser pulses\textcolor{blue}{,} one obtains a $\sqrt{2}$ improvement, again assuming Gaussian spectra \cite{lavoie09}.  It is this difference in resolution improvement---a factor of $\sqrt{2}$ for CPI vs $2$ for HOM---we address here.

Chirped pulses are a key resource in optical technologies, such as: chirped-pulse amplification \cite{strickland85}, optical pulse characterization \cite{iaconis98}, and single-photon bandwidth compression \cite{lavoie13}.  Most commonly, these techniques and demonstrations of CPI to date rely on 
\emph{linearly}-chirped pulses where the instantaneous frequency of light changes linearly in time.  Such pulses can be generated using configurations of standard optical components such as gratings or prisms \cite{treacy69}.  More complex pulses can be generated using ultrafast laser pulse-shaping techniques, such as those based on spatial light modulators \cite{weiner00, mazurek13}.  In this paper, we show that the CPI resolution can be enhanced when the laser pulses have a specific \emph{non}linear chirp to match, \emph{and even exceed}, the factor of $2$ enhancement from the quantum technique. In Sec.~\ref{sc:theory}, we present a theoretical description of chirped pulses and elements of interferometry.  In Sec.~\ref{sc:results}, we present numerical simulations of specific linearly and nonlinearly chirped pulses and the specific chirped pulse interferometry signals based on these pulses.

\section{Theory}\label{sc:theory}

\subsection{Chirped laser pulses}
\label{subsec:theorychirpedpulses}

A chirped pulse has an instantaneous frequency that changes in time \cite{diels06}.  They can be generated by starting with a transform-limited pulse with electric field, $E(\omega)$, with central frequency $\omega_0$ and applying a frequency-dependent phase, $\phi(\omega)$.  This phase implements a frequency-dependent group delay, $t(\omega) = \frac{d\phi}{d\omega}$.  

In the special case where the phase is \emph{quadratic} in the frequency, $\phi_L(\omega) = A(\omega-\omega_0)^2$, the group delay, $t(\omega)=2A(\omega-\omega_0)$, is \emph{linear} in the frequency.  Rearranging this expression, we have an expression for the instantaneous frequency $\omega(t) = \omega_0 + \frac{1}{2A} t$ which again is a linear relation.  We refer to such pulses as \emph{linearly chirped}.  Chirped pulses can also be generated by applying phases with more complicated frequency dependence.  Such pulses will, in general, have a nonlinear relation between the instantaneous frequency and time, so we refer to such pulses as \emph{nonlinearly chirped}.

Suppose an initially transform-limited laser pulse has a Gaussian spectrum,  
\begin{eqnarray} 
\label{eq:gaussianintensity}
I(\omega) &=& I_0 e^{-\frac{(\omega-\omega_0)^2\sigma_t^2}{4\ln2}},
\end{eqnarray}
where $\omega_0$ is the centre frequency, $\sigma_t$ is the full width at half maximum (FWHM) of the temporal intensity distribution, and $I_0$ is a constant.  Applying a quadratic phase maintains the Gaussian temporal intensity profile, but stretches the pulse in time to a width $\sigma_t\sqrt{1+ \left(8 \ln 2 \frac{A}{\sigma_t^2} \right)^2}$ (FWHM).  Linearly-chirped pulses maintain a Gaussian temporal profile so the intensity is higher in the centre where the instantaneous frequency is close to the centre frequency, $\omega_0$, and lower in the wings where the instantaneous frequency is far from the centre frequency.

\begin{figure}[t!]
\begin{center}
\includegraphics[width=0.85\columnwidth]{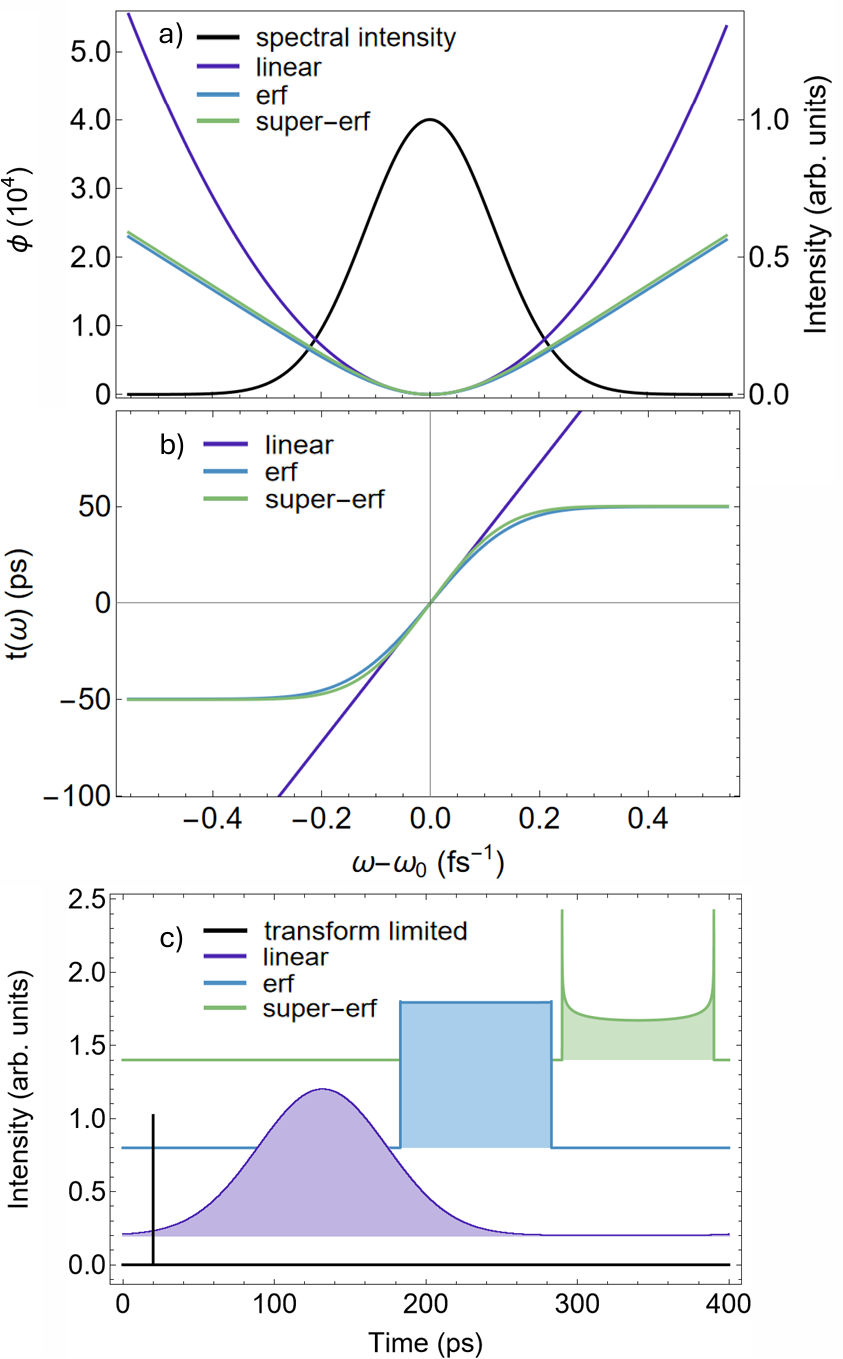}
\end{center}
\caption{
Spectral phase, time delay, and chirped-pulse intensity distributions.  A transform-limited Gaussian pulse with a temporal intensity distribution with width 10~fs (FWHM) (panel c), black line) and centre wavelength 800~nm has a Gaussian spectral intensity distribution (panel a), black line, right scale).
Applying the quadratic spectral phase $\phi_L(\omega)$ with $A=180,337$~fs$^{2}$ (panel a), purple line, left scale) generates a linear group delay as a function of frequency (panel b), purple line) and a linearly-chirped pulse with Gaussian temporal intensity profile (panel c), purple curve) with width 100~ps (FWHM).  Applying the erf-chirp spectral phase, $\phi_E(\omega)$, with $B=8300$ and $\sigma_t = 10$~fs (panel a), blue curve, left scale) to the initial pulse generates the group delay as a function of frequency (panel b), blue curve) and a nearly rectangular intensity profile, as expected, with width 100~ps (panel c), blue curve).  Applying the super-erf spectral phase, $\phi_{S}(\omega)$, with $B=7450$ and $\sigma_s = 11.2$~fs (panel a), green curve, left scale), generates the group delay as a function of frequency (panel b), green curve) and the temporal intensity distribution (panel c), green curve) with width 100~ps.  The group delay curves for the erf and super-erf-chirped pulses show that many frequencies will be delayed by close to $-50$~ps or $+50$~ps.  The erf and super-erf phase and group delay functions are very similar despite the significantly different temporal intensity distributions.  Note: for clarity, the pulses in panel c) have been shifted temporally so they do not overlap in the figure. 
\label{fig:pulseshapefigure}}
\end{figure}

We aim to find a phase function, $\phi_{E}(\omega)$, that stretches a transform-limited pulse with the intensity distribution, $I(\omega)$, to a nonlinearly-chirped pulse with a (nearly) constant intensity, i.e., a rectangular pulse.  We can express this condition as $dt/d\omega \propto I(\omega)$, or $t(\omega) \propto \int I(\omega) d\omega$, up to an integration constant.  The spectral phase required to implement this frequency-dependent temporal shift is $\phi_{E}(\omega) = \int t(\omega) d\omega$ \cite{mazurek13b},
\begin{eqnarray}
\phi_E(\omega) &=& B\left(\frac{e^{-x^2}-1}{\sqrt{\pi}}+x\hspace{0.5mm}\mathrm{erf}{x}\right)
\end{eqnarray}
where $x=(\omega-\omega_0)\sigma_t/\sqrt{4\ln2}$, $B$ is a constant, and the integration constant was chosen to set $\phi(\omega_0) = 0$.
We refer to this particular nonlinearly-chirped pulse as an \emph{erf-chirped} pulse. This is a complicated looking expression, however its behaviour is straightforward; it is approximately quadratic near the minimum and transitions toward more linear growth at a characteristic width controlled by the parameter $\sigma_t$.  An example is shown in Fig.~\ref{fig:pulseshapefigure}a) (blue curve, left scale).  Note that this particular phase function was derived assuming a Gaussian spectrum; for a different spectral profile, an alternate phase function would be required to generate a rectangular temporal profile.

We derived the phase, $\phi_{E}(\omega)$, with the purpose of generating a chirped pulse with a rectangular intensity distribution.  However, we can take the same functional form and define a family of phase functions,
\begin{eqnarray}
\phi_{S}(\omega) &=& B\left(\frac{e^{-y^2}-1}{\sqrt{\pi}}+y\hspace{0.5mm}\mathrm{erf}{y}\right)
\end{eqnarray}
where $y = (\omega-\omega_0)\sigma_{s}/\sqrt{4\ln 2}$, where $\sigma_s>\sigma_t$.  As we will see, applying such spectral phases to a Gaussian transform limited pulse will overstretch the centre of the pulse relative to the wings, leading to a double-peaked pulse.  We refer to these pulses as \emph{super-erf} chirped.  An example is shown in Fig.~\ref{fig:pulseshapefigure}a) (green curve, left scale).

\begin{figure*}[t]
\begin{center}
\includegraphics[width=2\columnwidth]{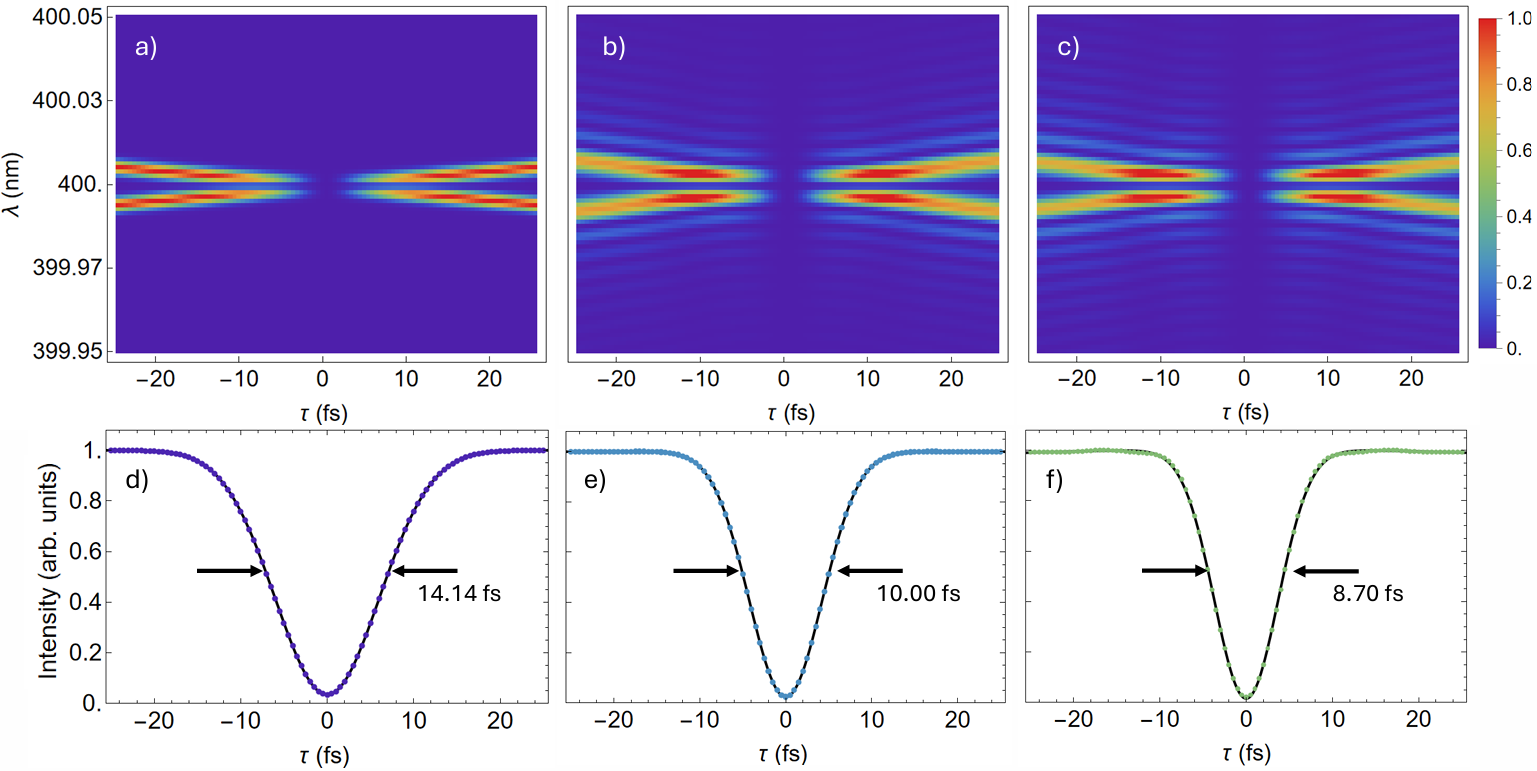}
\end{center}
\caption{Chirped-pulse interferometry with different chirped pulses.  The top set of panels a)-c) show the CPI signal spectral intensity in the absence of unbalanced dispersion as a function of delay using pairs of a) linearly, b) erf, and c) super-erf-chirped pulses. To generate the bottom set of panels d)-f), we integrated the SFG signal over $1$~nm wavelength range, centred on $400$~nm (twice the centre frequency of the initial pulse) to give the signal intensity as a function of delay, $\tau$.  In each case, the signal has a high-visibility dip in the intensity at zero delay.  The widths of the dips are 14.14~fs, 10.00~fs, and 8.70~fs (FWHM) for the d) linear, e) erf, and f) super-erf-chirped pulses, respectively.  A white-light interferometer (WLI) signal would be expected to have an envelope with width 20~fs (twice the initial transform pulse duration) (FWHM).  The CPI signals show improvement in resolution over WLI by a factor of $1.41\approx\sqrt{2}$, $2.00$, and $2.30$ for the linear, erf, and super-erf-chirped pulses, respectively.  
\label{fig:xplotsanddips}}
\end{figure*}

\subsection{Chirped-pulse interferometry}
\label{subsec:theorycpi}

The CPI signal from the setup shown in Fig.~\ref{fig:interferometersfigure}~c) can be calculated through the approach in Ref.~\cite{resch09}.  A pair of synchronous, oppositely-chirped pulses with electric fields written in the frequency domain $E_A(\omega) = E(\omega)\exp(-i\phi(\omega))$ and $E_C(\omega)=E(\omega)\exp(i\phi(\omega))$ are first combined on a beamsplitter, generating two new fields $E_1(\omega) \propto E_A(\omega) + E_C(\omega)$ and $E_2(\omega) \propto E_A(\omega)-E_C(\omega)$ one in each arm of the interferometer.  We can add unbalanced quadratic dispersion characterized by a parameter, $\epsilon$, a time delay, $\tau$, to the pulses of the fields through the transformations $E_1(\omega, \epsilon) = E_1(\omega)\exp{i \epsilon(\omega-\omega_0)^2}$ and $E_2(\omega, \tau) = E_2(\omega)\exp{i\omega \tau}$.  These fields drive sum-frequency generation (SFG), which we approximate using the expression $E_{SFG}(t,\tau,\epsilon) \sim  E_1(t, \epsilon)E_2(t,\tau)$, where the fields are now expressed in the time domain.  The output SFG signal as a function of frequency, delay, and dispersion is then,
\begin{eqnarray}\label{eq:cpisignal}
I(\omega,\tau,\epsilon) \propto \Big| \mathscr{F} \bigl[E_1(t,\epsilon)E_2(t,\tau)\bigl] \Big|^2
\end{eqnarray}
where $\mathscr{F}$ is the Fourier transform. We then apply a spectral filter over a narrow frequency range near twice the centre frequency of the chirped pulses and measure the resulting signal intensity as a function of the delay.

If the chirped pulses used in CPI are linearly-chirped Gaussian pulses, then the nonlinear SFG signal in CPI overweights contributions from the frequencies in the centre of the pulse and underweights contributions from the frequencies in the wings.  This lowers the effective bandwidth generating the signal, lowering resolution.  It suggests the strategy of using nonlinearly-chirped pulses, specifically the erf and super-erf-chirped pulses, may not suffer from this effective bandwidth reduction and exhibit higher resolution.  

CPI with linearly-chirped pulses has been shown to be robust against unbalanced quadratic dispersion.  It operates in the regime where $A\gg\epsilon$, i.e., the chirp on the laser pulses, $A$, is much larger than the unbalanced dispersion, $\epsilon$.  This reduces the impact of dispersion on the CPI signal by a factor of $\sim\epsilon/A$ \cite{resch09}.  In practice, one may select the value of $A$ based on a maximum value of $\epsilon$ one expects to encounter in a particular application. Stretching the pulses more than is required to effectively cancel the maximum expected dispersion further decreases the peak intensity and lowers the CPI signal generation efficiency.  Moreover, since CPI only cancels even orders of dispersion, it offers little advantage to keep increasing the chirp once the odd orders of dispersion become significant.

\subsection{White light interferometry}
In contrast to CPI, WLI is sensitive to all orders of unbalanced dispersion.  We model the measured signal as a function of delay using, $S(\tau) \propto \int d\omega I(\omega) \left[1+\cos(\omega \tau - \epsilon(\omega-\omega_0)^2)\right]$, with the spectrum $I(\omega)$ from Eq.~\eqref{eq:gaussianintensity}.  This signal has wavelength-scale fringes modulated by a Gaussian envelope with width $\sigma_t(\epsilon) = 2\sigma_t \sqrt{1 + \left(4\ln{2} \frac{\epsilon}{\sigma_t^2}\right)^2}$ (FWHM).  For zero dispersion, this envelope has twice the temporal width of the pulse and the width grows rapidly with the dispersion, $\epsilon$.

\section{Results and discussion}\label{sc:results}

\subsection{Numerically modelling chirped pulses}
\label{subsec:resultschirpedpulses}

We begin with a Gaussian transform-limited pulse with a temporal intensity distribution $10$~fs wide (FWHM) and centre wavelength 800~nm (or $\omega_0 \approx 2.35$~fs$^{-1}$).  The spectral intensity distribution is shown in Fig.~\ref{fig:pulseshapefigure}a) (black line, right scales) and the normalized intensity temporal distribution in Fig.~\ref{fig:pulseshapefigure}c) (black line).  We consider the action of the linear, erf, and super-erf chirps on this pulse by applying a Fourier transform to the initial pulse to bring it to the frequency domain, applying the phase, then applying the inverse Fourier transform back to the time domain.  We sample the signal with $N=2^{19}$ equally spaced data points over a time range of $400$~ps.  Calculations are performed with the discrete Fourier and inverse Fourier transform functions in Mathematica.  We produce chirped pulses with temporal intensity distributions of $100$~ps wide with each of the linear, erf, and super-erf phase functions.  Specifically we choose the parameter $A=180,337$~fs$^{2}$ for the linearly-chirped pulse; $B = 8300$ and $\sigma_t = 10$~fs (matching the width of the initial pulse) for the erf-chirped pulse; and $B = 7450$ and $\sigma_s = 11.2$~fs for the super-erf-chirped pulse.  The spectral phases, group delays, and temporal intensity distributions for the chirped pulses are shown in Fig.~\ref{fig:pulseshapefigure}.  In Fig.~\ref{fig:pulseshapefigure}c) we see the linearly-chirped pulse profile (purple) maintains a Gaussian shape; the erf-chirped pulse profile (blue) is approximately rectangular as expected; and the super-erf-chirped pulse (green) has a double-peaked shape with the intensity in the centre of the pulse lower than that in the leading and lagging edges.  Note that in all cases the spectral intensity distributions of the pulses remain Gaussian as it is unchanged by a purely dispersive phase.  

\subsection{Chirped-pulse interferometry with no dispersion}
\label{subsec:resultscpi}

We use Eq.~\eqref{eq:cpisignal} to calculate the CPI signal, $I(\omega,\tau,\epsilon)$, using oppositely chirped pulse pairs of the three types described in the previous section.  As before we take $N=2^{19}$ evenly sampled points over a time range of $400$~ps.  We calculate the signal over a range of delays from $-25$~fs to $+25$~fs in $0.5$~fs steps.

We start by setting the unbalanced dispersion to zero and calculate $I(\omega,\tau,\epsilon=0)$ in narrow wavelength range near 400~nm (i.e., twice the centre frequency of the initial pulse) as a function of delay.  We express this spectral intensity distribution in terms of wavelength as a function of the delay and plot it in false color in the top three panels of Fig.~\ref{fig:xplotsanddips} for the a) linearly, b) erf, and c) super-erf-chirped pulses using pulse parameters from Sec.~\ref{subsec:resultschirpedpulses}.  To first approximation, at large delays the SFG signal consists of a bright doublet of narrow spectral lines and as the delay gets smaller, the spacing between the lines decreases. These lines arise from the SFG generated from pairs of oppositely chirped pulses with a relative time delay.  Near zero delay, the lines overlap and interfere destructively eliminating the intensity at zero delay.  The erf and super-erf pulses exhibit additional spectral features.  These stem from the fact that the sum of the instantaneous frequencies from a pair of nonlinearly and oppositely chirped pulses with a relative time delay is not constant along the length of the pulses, as it is in the linear case.

To obtain the CPI interference dip, closely analogous to that generated in the quantum Hong-Ou-Mandel interferometer, we sum
the signal, $I(\omega,\tau,\epsilon=0)$, over a 1~nm range about the centre wavelength of 400~nm.  We plot this integrated signal intensity vs delay in the lower three panels of Fig.~\ref{fig:xplotsanddips} for the d) linear, e) erf, and f) super-erf-chirped pulses.  In each case, we see a high-visibility interference dip in the intensity at zero delay. The widths of the dips are 14.14, 10.00, and 8.70~fs (FWHM) for the d) linear, e) erf, and f) super-erf-chirped pulses, respectively.  Recall that the initial pulse has
a temporal intensity width 10~fs FWHM.  The white-light interference pattern from such a pulse with no unbalanced dispersion would be twice as broad, 20~fs FWHM.  Thus the CPI signal resolution is enhanced by factors of d) $1.41\approx \sqrt{2}$, e) $2.00$, and f) $2.30$ for the linear, erf, and super-erf-chirped pulses, respectively.

\subsection{Chirped-pulse interferometry with unbalanced dispersion}
\label{subsec:resultsdispcanc}

\begin{figure*}[t]
\begin{center}
\includegraphics[width=1.75\columnwidth]{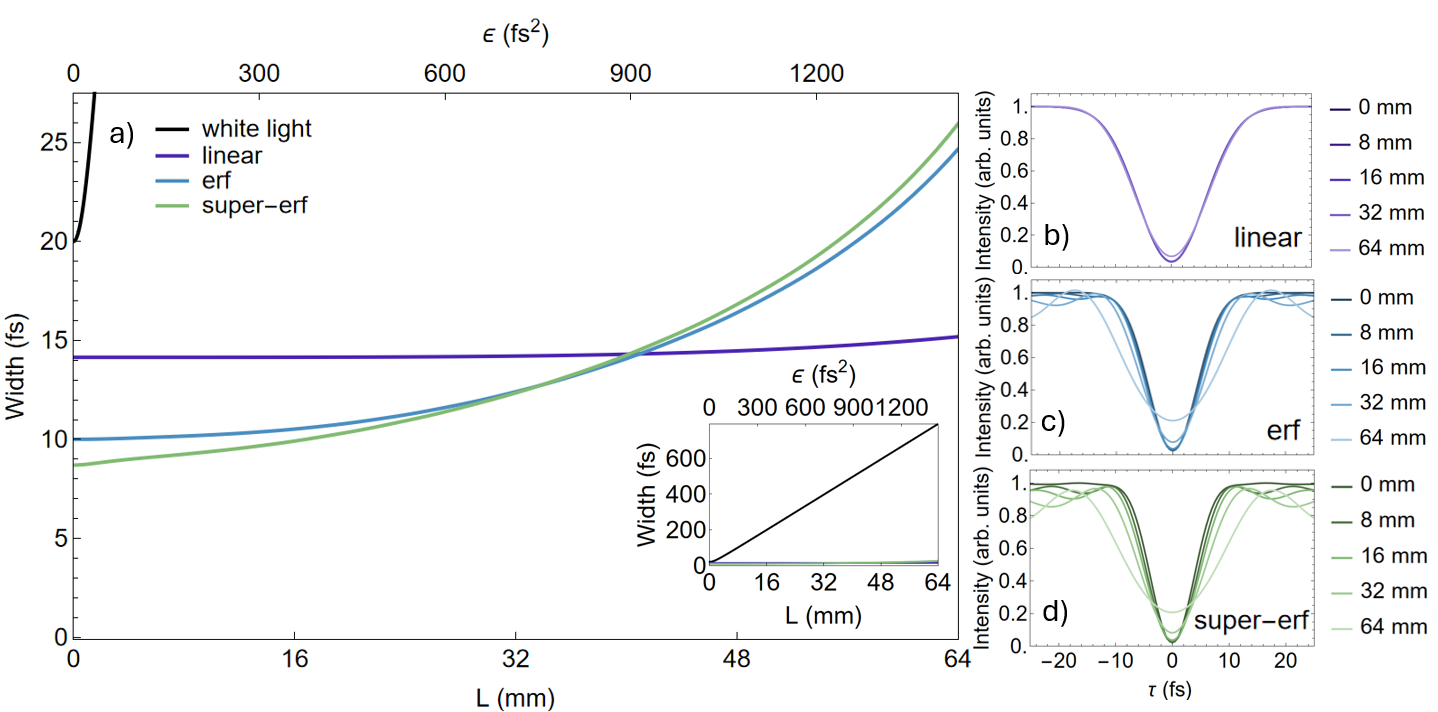} 
\end{center}
\caption{CPI signal widths vs dispersion.  In panel a), the WLI and CPI interference signal widths are shown as a function of unbalanced dispersion, $\epsilon$. The CPI signal, $I(\omega,\tau,\epsilon)$, is
summed over a 1~nm range about 400~nm to obtain the CPI interference dips using b) linearly, c) erf, and d) super-erf-chirped pulses for quadratic dispersion strength, $\epsilon$, corresponding to a length (L) from 0~mm to 64~mm of BK7.  In panel b), the linearly-chirped CPI interference patterns show almost no change due to dispersion, with a slight drop in visibility and slightly increased width at the highest dispersion values.  Panels c) and d) show that the erf and super-erf CPI interference patterns are more impacted by dispersion exhibiting some broadening and distortion, most evident for the largest dispersion values.  In panel a), the CPI interference dips generated by linearly-chirped pulses have
almost no dependence on the dispersion, indicating near perfect dispersion cancellation.  The CPI interference dips generated by the erf and super-erf-chirped pulses are narrower than the dips generated by the linearly-chirped pulses, but are more impacted by dispersion.  The white-light interference signal starts at 20~fs,  grows quickly 
with dispersion, rapidly going off the scale; the inset in panel a) shows the significant and nearly linear increase in WLI signal width with dispersion.
\label{fig:dispersion}}
\end{figure*}

To investigate dispersion cancellation in the case of the nonlinearly-chirped pulses, we model unbalanced second-order dispersion as described in Sec.~\ref{subsec:theorycpi}.  Using the Sellmeier coefficients for the common BK7 glass from Ref.~\cite{schott14}\textcolor{blue}{,} we calculated the second-order dispersion coefficient $\epsilon/\mathrm{mm} = \frac{1}{2} d^2 k/d\omega^2 = 22.3238~$fs$^2$/mm at $800~$nm.  We vary the amount of unbalanced dispersion in our calculations by including the quadratic dispersion from different thicknesses of BK7 in one interferometer arm.

We calculate
the CPI signal as a function of delay as described in Secs.~\ref{subsec:theorycpi} and \ref{subsec:resultscpi}, by first calculating $I(\omega,\tau, \epsilon)$ for values of epsilon ranging from $0$~mm to $64$~mm of BK7, then summing the intensities over a 1~nm range.  Example calculated CPI signals are shown in the right three panels of  Fig.~\ref{fig:dispersion} for the b) linear, c) erf, and d) super-erf-chirped pulses.  These figures show minimal effect of dispersion for the linearly-chirped pulses.  The erf and super-erf-chirped pulses show some broadening and distortion, in the form of ringing, for the largest amounts of dispersion.

To summarize the effect of these, and other intermediate values of dispersion, we fit the interference dips using a Gaussian function.  The results for the widths (FWHM) are shown in Fig.~\ref{fig:dispersion}a).  For $0$~mm dispersion, the widths of the interference dips are the same as those reported in Sec~\ref{subsec:resultscpi}.  As we increase the dispersion, the width of the linearly-chirped CPI signal shows almost no change with a small amount of broadening apparent for the largest amounts of dispersion; the signal broadens from 14.14~fs with zero dispersion to 15.18~fs with 64~mm of BK7, just a 7\% increase.  The CPI signal from the erf and super-erf-chirped pulses begin with a narrower interference dip but exhibit more sensitivity to dispersion than the linearly-chirped pulse.  The erf and super-erf CPI widths grow from 10.00~fs and 8.70~fs with zero dispersion to 24.66~fs and 25.94~fs respectively with 64~mm of BK7, corresponding to 147\% and 198\% increases.  For context, we show the width of the envelope of a white-light interference pattern generated by light with the same bandwidth of the chirped pulses in Fig.~\ref{fig:dispersion}a) and the inset as a function of $\epsilon$.  Here we see nearly linear growth in the envelope with the WLI signal width
broadening to 792.51~fs at 64~mm of BK7, a total increase of $\sim$3900\%.

CPI with erf and super-erf-chirped pulses exhibit a significant reduction in sensitivity to dispersion compared to white-light interferometry, but are somewhat more sensitive to dispersion than CPI with linearly-chirped pulses.  There are two factors contributing to this difference.  First, for CPI with linearly-chirped pulses, effective dispersion is lowered by a factor of $\epsilon/A$~\cite{resch09}.  With erf and super-erf-chirped pulses, the pulses are stretched by a large factor in the centre and comparatively less in the wings.  As a result, it is reasonable to expect the parts of the signal generated by frequencies in the middle will be less sensitive to dispersion and those generated by frequencies in the wings to be more sensitive.  Second, we compare signals from different types of chirped pulses with the same width.  It is apparent from Fig.~\ref{fig:pulseshapefigure}a) and b) that the maximum slope of the phase and maximum time delay is larger for the linearly-chirped pulses than the erf and super-erf-chirped pulses under these conditions.   Comparing pulses of equal temporal width may not be a fair comparison, giving the linearly-chirped case an advantage for dispersion cancellation.  It is certainly more demanding experimentally to implement large time delays; for example a pulse shaper using a spatial-light modulator with a fixed number of pixels can only faithfully implement spectral phase functions up to a maximum slope determined by the Nyquist limit. Despite these disadvantages, CPI with erf and super-erf-chirped pulses cancel a significant amount of quadratic dispersion and outperform the resolution of CPI with linearly-chirped pulses up to about 40~mm of BK7 with the parameters we considered; this particular value of dispersion could be pushed higher by increasing the chirping strength.

\section{Conclusions}

Chirped-pulse interferometry is a classical white-light interferometry technique that shares important features with the quantum Hong-Ou-Mandel interferometer using energy-time entangled photon pairs.  These include a phase insensitive, high visibility interference dip and automatic dispersion cancellation.  

In the absence of dispersion, CPI based on linearly-chirped Gaussian pulses exhibits an interference dip that is $\sqrt{2}$ narrower than the white-light signal generated by one of the pulses.  If we instead use erf-chirped pulses which generate nearly rectangular pulses with a Gaussian spectrum, the resolution is enhanced to a factor of 2 narrower than the white-light signal, matching the advantage from the quantum interferometer.  If we use super-erf-chirped pulses, as described in the paper, this advantage can exceed the factor of 2 to about 2.3.  Our results show that CPI signals with linear, erf, and super-erf-chirped pulses are robust against unbalanced dispersion, compared to white-light interferometry.  However, CPI with erf and super-erf-chirped pulses are more sensitive to dispersion than CPI with linearly-chirped pulses, so there is a tradeoff in obtaining enhanced resolution.

In this work, we have evaluated the different chirping functions based on the widths of the dips produced. We note that in practice, signal intensity is also important to consider. Signal intensity in CPI is affected by both the temporal shape and width of the chirped pulses; in general, stretching a pulse more will decrease signal. In this work, for the specific parameters we chose for our simulations, the signals generated by the erf and super-erf pulses were approximately 1.5 times more intense than those generated by the linear chirp in the zero dispersion case. While this provides an additional benefit for the nonlinear chirping functions here, 
the difference stems from our decision to compare pulses with the same FWHM and would not hold if we had fixed some other property, e.g. maximum time delay.

This work raises several interesting questions for further study.  How narrow can the CPI interference dip be using pulses in the super-erf-chirped class of pulses, or more generally, using any nonlinear chirping function?  What are the relationships between strength of chirp, resolution, dispersion cancellation, and signal generation efficiency? Do linearly-chirped pulses maintain their advantage in dispersion cancellation once third-order dispersion in realistic materials is considered?  How robust are these methods once experimental constraints on pulse-shaping components are considered \cite{weiner00,mazurek13}?  How will the erf and super-erf-chirped CPI perform when the beam in one arm reflects from a material with multiple interfaces, as is important for dispersion-cancelled OCT imaging \cite{abouraddy02, nasr03, lavoie09, mazurek13}?  

We have shown how a family of nonlinearly-chirped optical pulses can enhance the performance of chirped-pulse interferometry to meet and even exceed that of the analogous quantum device.  Nonlinearly-chirped pulses open a new range of exciting possibilities for interferometry.

\section{Acknowledgments}
The authors thank Rajibul Islam for interesting discussions.  This research was
supported in part by the Natural Sciences and Engineering
Research Council of Canada (NSERC), Industry Canada, the Canada Foundation for Innovation
(CFI), and the Canada First Research Excellence Fund
(CFREF).

\end{document}